\documentstyle[aps,prl,multicol]{revtex}
\begin{document}
\title{\bf Spatial Persistence of Fluctuating Interfaces}

\author{Satya N. Majumdar$^{(1),(2)}$ and Alan J. Bray$^{(3)}$}

\address{(1)Laboratoire de Physique Quantique (UMR C5626 du CNRS), 
Universit\'e Paul Sabatier, 31062 Toulouse Cedex, France. \\
(2) Tata Institute of Fundamental Research, Homi Bhabha Road, 
Mumbai-400005, India. \\
(3)Department of Physics and Astronomy, University of Manchester,  
Manchester, M13 9PL, UK}

\date{\today}

\maketitle

\begin{abstract}
We  show  that  the  probability,  $P_0(l)$,  that  the  height  of  a
fluctuating  $(d+1)$-dimensional interface in  its steady  state stays
above its initial value up to  a distance $l$, along any linear cut in
the  $d$-dimensional space, decays  as $P_0(l)\sim  l^{-\theta}$. Here
$\theta$  is a  `spatial'  persistence exponent,  and takes  different
values,  $\theta_s$ or  $\theta_0$, depending  on how  the  point from
which $l$  is measured  is specified. While  $\theta_s$ is  related to
fractional Brownian motion, and  can be determined exactly, $\theta_0$
is non-trivial even for Gaussian interfaces.

\noindent

\end{abstract}

\begin{multicols}{2}

Fluctuating  interfaces  have  played   the  role  of  a  paradigm  in
nonequilibrium  statistical physics  for  the last  two decades,  with
applications   ranging   from    growth   problems   to   fluid   flow
\cite{Krug1}. Traditionally the  different universality classes of the
interfaces are specified by  the exponents associated with the dynamic
correlation      functions.      Recently      the      concept     of
persistence\cite{review},   i.e.\  the  statistics   of  first-passage
events,  was used  to characterize  the {\em  temporal} history  of an
evolving interface\cite{Krug2}. Starting  from an initially {\em flat}
interface, it was  found that the probability that  the interface at a
given point  in space stays  above its initial  height up to  time $t$
decays as a  power-law in time with an  exponent $\theta_0^{tem}$ that
is  nontrivial  even  for  simple `Gaussian'  interfaces  (i.e.\  with
dynamics     governed    by     a     linear    Langevin     equation)
\cite{Krug2,Kallabis}.  On  the other hand, the  statistics of returns
to  an  initial  profile  chosen  from  the  ensemble  of  equilibrium
configurations  was  shown to  be  governed  by  a different  exponent
$\theta_s^{tem}$.   The  exponent $\theta_s^{tem}$  was  shown be  the
identical  to the first  passage exponent  of the  fractional Brownian
motion  (fbm)  which is  known  exactly\cite{Krug2}.  In  \cite{Krug2}
these  two   regimes  were  termed  the  {\em   coarsening}  and  {\em
stationary} regimes respectively.

The   exponents   $\theta_0^{tem}$   and   $\theta_s^{tem}$   describe
first-passage  properties in  {\em time}.   A question  that naturally
arises is: what about the  first-passage properties in space? Is there
a power-law  distribution and associated nontrivial  exponents for the
{\em spatial} persistence  of an interface?  In this  paper we address
this important question and show that indeed in the {\em steady state}
(long time  limit), the probability  $P_0(l)$ that an  interface stays
above its initial value over a distance $l$ from a given point in {\em
space} decays as $P_0(l) \sim l^{-\theta}$ for large $l$. Furthermore,
analogous  to  its   temporal  counterpart,  the  spatial  persistence
exponent  $\theta$  again   takes  different  values,  $\theta_0$  and
$\theta_s$,  according  to the  ``initial''  condition,  in this  case
specified by the height and  its spatial derivatives at the point (the
``initial point'' in space) from which $l$ is measured. If the initial
point  is  sampled  uniformly   from  the  ensemble  of  steady  state
configurations, the  relevant exponent, $\theta_s$, is  related to the
Hurst exponent  of the spatial roughening. Conversely,  if the initial
point is such that the  height and its spatial derivatives are finite,
independent  of the  system size  (we will  call this  a  {\em finite}
initial condition),  the corresponding  exponent $\theta_0$ is  a new,
nontrivial exponent.

A  second   interesting  question  asks:  Is   there  a  morphological
transition in the stationary profile of a fluctuating interface if one
changes the  mechanism of fluctuations by tuning  either the dynamical
exponent $z$ or the spatial dimension $d$? This question has important
experimental  significance.  For  example,  recent Scanning  Tunneling
Microscope  (STM)   measurements  have  shown\cite{Giesen}   that  the
dynamical exponent $z$ characterizing the fluctuations of single layer
Cu(111) surface  changes from $z=2$  at high temperatures to  $z=4$ at
low   temperatures\cite{Giesen,TW,KE,Toro}.    Thus   by  tuning   the
temperature, and hence $z$, the STM measurements may possibly detect a
morphological transition if there is  one.  We show here that Gaussian
interfaces indeed exhibit a morphological transition across a critical
line $z_c(d)=d+2$ in the $(z-d)$ plane as one tunes either $z$ or $d$.
For  $z>z_c(d)$, the  steady state  profile  is smooth  with a  finite
density of zero crossings along  any linear cut in the $d$-dimensional
space. On the contrary, for  $z<z_c(d)$, the density of zero crossings
is  infinite and  the locations  of zeros  are nonuniform  and  form a
fractal set.

A seemingly unrelated problem is the usual temporal persistence of the
stochastic process\cite{diff},
\begin{equation}
{ {d^nx}\over {dt^n}}=\eta(t),
\label{st1}
\end{equation}
where $x$  is the position of  a particle and $\eta(t)$  is a Gaussian
white   noise  with   zero  mean   and  delta   correlation,  $\langle
\eta(t)\eta(t')\rangle =\delta(t-t')$. The probability, $P_0(t)$, that
$x$  does  not  change sign  up  to  time  $t$ decays  as  $P_0(t)\sim
t^{-\theta(n)}$  for   large  $t$,  where   the  persistence  exponent
$\theta(n)$ depends continuously  on $n$\cite{review}. The exact value
of the exponent  is known only for $n=1$  (the usual Brownian motion),
$\theta(1)=1/2$\cite{Feller},  and   $n=2$  (the  random  acceleration
problem),  $\theta(2)=1/4$\cite{TB1,SGL}.    The  latter  problem  has
recently  attracted  attention  in  connection  with  the  problem  of
inelastic collapse in granular materials\cite{CSB}.

In this paper,  we establish an intricate mapping  between the process
in  Eq.\  (\ref{st1})  and   the  steady  state  profile  of  Gaussian
interfaces evolving via the Langevin equation,
\begin{equation}
{{\partial  h}\over {\partial  t}} =  - {\large  (-{\nabla}^2\large
)}^{z/2}h + \xi
\label{lang}
\end{equation} 
where ${\nabla}^2$  is the $d$-dimensional Laplacian  operator, $z$ is
the dynamical exponent associated  with the interface and $\xi(\vec r,
t)$ is  a Gaussian  white noise with  zero mean and  $\langle \xi(\vec
r,t)\xi   ({\vec  r}',   t')\rangle=2\delta(\vec   r-{\vec  r}')\delta
(t-t')$.  The continuum equation (\ref{lang}) is well defined only for
$z>d$. For $z<d$,  one needs a nonzero lattice  constant in real space
or an  ultraviolet cut-off  in momentum space.   Below we  will assume
$z>d$.

Let us  summarize our  main results. We  find two  independent spatial
persistence  exponents  $\theta_0$  and  $\theta_s$ depending  on  the
initial point in space (from  where measurements start).  The two main
results are: (i)  By exploiting the mapping to  the stochastic process
(\ref{st1}), we show that for a `finite' initial starting point,
\begin{equation}
\theta_0= \theta(n), \ {\rm where}\ n=(z-d+1)/2.
\label{n}
\end{equation}
and   (ii)  For  initial   points  sampled   from  the   steady  state
configurations, however, we show exactly that
\begin{equation}
\theta_s=\cases{3/2-n,\,  &$1/2<n<3/2$ \cr
         0,\, &$n>3/2$.   \cr}
\label{steady}
\end{equation}
where $n=(z-d+1)/2$.
                                
Exploiting the two  exact results, $\theta(1)=1/2$ and $\theta(2)=1/4$
in Eq.\ (\ref{n}), we  find that for any $(d+1)$-dimensional interface
with $z-d=1$, $\theta_0=1/2$  and for $z-d=3$, $\theta_0=1/4$ exactly.
An explicit example  of the former case is  the $(1+1)$-dimensional EW
model  with  $z=2$ and  $d=1$,  giving  $\theta_0=1/2$ \cite{DB}.   An
example  of  the  latter  case is  the  $(1+1)$-dimensional  continuum
version of the Das-Sarma Tamborenea (DT) model\cite{DT} with $z=4$ and
$d=1$, which  gives $\theta_0=1/4$, a  new exact result.   For general
$n>3/2$, we show  that the exponent $\theta (n)$  and hence $\theta_0$
can be  estimated rather accurately by using  an `independent interval
approximation' (IIA).   Thus Eq.\ (\ref{n}),  besides giving accurate,
and in some cases exact, results for the new exponent $\theta_0$, also
give a physical meaning to  the stochastic process in Eq.\ (\ref{st1})
for general $n$.  We will use this correspondence to further establish
a  morphological  transition   for  Gaussian  interfaces,  across  the
critical line  $z_c(d)=d+2$, as  one tunes $z$  or $d$ in  the $(z-d)$
plane.

Let  us consider  the $n$-th  derivative of  the scalar  field $h(\vec
r,t)$  with respect  to any  particular direction,  say $x_1$,  in the
$d$-dimensional  space, $g(\vec r,  t)= \partial^n  h/\partial x_1^n$.
From  Eq.\  (\ref{lang}),  it  follows  that  the  Fourier  transform,
${\tilde  g}(\vec   k,t)=\int  g(\vec  r,t)e^{i{\vec   k}{\bf  .}{\vec
r}}d^d{\vec r}$ then evolves as,
\begin{equation}
{  {\partial {\tilde g}(\vec  k,t)}\over {\partial  t}}=-|k|^z {\tilde
g}(\vec k, t) + \xi_1(\vec k,t)
\label{lang1}
\end{equation}
where $|k|^2=k_1^2+k_2^2+\ldots$ and $\xi_1$  is a Gaussian noise with
zero   mean  and   correlator,  $\langle   \xi_1(\vec  k,t)\xi_1({\vec
k}',t')\rangle=2  |k_1|^{2n}  \delta(\vec  k+{\vec  k}')\delta(t-t')$.
Using  Eq.\ (\ref{lang1}),  one  can easily  evaluate the  correlator,
$\langle  g(\vec  k,  t)g({\vec  k}',t)\rangle=G(\vec  k,t)\delta(\vec
k+{\vec  k}')$,   to  obtain,  in   the  steady  state,   $G(\vec  k)=
|k_1|^{2n}/|k|^z$.   Inverting  the   Fourier   transform  gives   the
stationary, real-space correlator,
\begin{equation}
\langle     g(x_1,    x_2,\ldots)g({x'}_1,    x_2,\ldots)\rangle=\int{
{|k_1|^{2n}}\over {|k|^z} }e^{ik_1(x_1-{x'}_1)} d^d{\vec k}.
\label{corr2}
\end{equation}
It is  now easy to  see from Eq.\  (\ref{corr2}) that with  the choice
$n=(z-d+1)/2$ the integral yields a delta function, i.e.,
\begin{equation}
\langle g(x_1,x_2,x_3,\ldots)g({x'}_1,x_2,x_3,\ldots)\rangle
=D\delta(x_1-{x'}_1)
\label{corr3}
\end{equation}
where $D$ is just a dimension-dependent number.

Since  the basic  Langevin equation,  Eq.\ (\ref{lang}),  is  a linear
equation and  the noise $\xi$ is  Gaussian, clearly $h$  is a Gaussian
process and therefore  its $n$-th derivative, $g$, is  also a Gaussian
process. A  Gaussian process is completely specified  by its two-point
correlator. Hence  in the  steady state limit  $t\to \infty$,  one can
write   the   following    effective   equation,   which   gives   the
$x_1$-dependence of $g$ at fixed $x_2,x_3,\ldots$
\begin{equation}
g(x_1,x_2,\ldots)={{{\partial}^nh}\over{{\partial x_1}^n}}=\eta(x_1),
\label{eff1}
\end{equation}  
where  $\eta$ is  Gaussian white  noise  with zero  mean and  $\langle
\eta(x_1)\eta({x'}_1)\rangle=\delta(x_1-{x'}_1)$.   Note that  we have
rescaled the  $x_1$ axis to  absorb the constant $D$.   This effective
stochastic equation will generate  the correct two-point correlator as
given by Eq.\ (\ref{corr3}) and  also all higher order correlations as
well  since $g$  is a  Gaussian process.   Making  the identification,
$x_1\Longleftrightarrow   t$   and   $h\Longleftrightarrow  x$,   Eq.\
(\ref{eff1}) immediately  reduces to the process  in Eq.\ (\ref{st1}).
Hence the  steady state profile  of a Gaussian interface  evolving via
Eq.\  (\ref{lang})  can be  effectively  described  by the  stochastic
equation  (\ref{st1})  with   $n=(z-d+1)/2$.  Therefore,  the  spatial
persistence  of  the Gaussian  interface  also  gets  mapped onto  the
temporal persistence of the stochastic process (\ref{st1}).

Consider the stochastic process  (\ref{st1}) starting from the initial
condition at $t=0$ where $x$ and  all its time derivatives up to order
$(n-1)$  vanish. It  turns out  that the  statistics of  first passage
events  of  this  process  depends  crucially on  whether  one  starts
measuring these events  right from $t=0$ or or if  one first waits for
an infinite time and then starts measuring the events. This is similar
to  the  `coarsening'  versus   `stationary'  regimes  as  studied  in
\cite{Krug2}.  This  can be  quantified  more  precisely  in terms  of
two-time correlations.  Integrating Eq.\ (1) $n$ times gives
\begin{equation}
x(t)={1\over {\Gamma (n)}}\int_0^t\eta(t_1)(t-t_1)^{n-1}dt_1,
\label{lap2}
\end{equation}
where   $\Gamma(n)$  is   the   usual  Gamma   function.   From   Eq.\
(\ref{lap2}), which  provides a natural continuation  from integer $n$
to real $n$, it is easy to evaluate the correlation function,
\begin{eqnarray}
C(t,t',t_0)&=&\langle [x(t_0 + t) - x(t_0)]
[x(t_0 + t')- x(t_0)]\rangle \nonumber \\
&=&A(t_0+t,t_0+t')-A(t_0+t,t_0) \nonumber \\
&-&A(t_0,t_0+t')+A(t_0,t_0),
\label{hyper}
\end{eqnarray}
where the autocorrelation $A(t,t')$ is given by,
\begin{equation}
A(t,t')={1\over{\Gamma^2(n)}}\int_0^{min(t,t')}
(t-t_1)^{n-1}(t'-t_1)^{n-1}dt_1.
\label{lap3}
\end{equation}             
Thus the  correlation function  that fully characterizes  the Gaussian
process  and  thereby  its  persistence,  depends  explicitly  on  the
`waiting'  time $t_0$, the  starting point  of measurements.  It turns
out, however, that there  are only two asymptotic behaviors controlled
respectively  by the  $t_0=0$ and  $t_0\to \infty$  fixed  points. Any
finite $t_0$ flows into the $t_0=0$ fixed point.

To  see how  the  two exponents,  $\theta_0$  and $\theta_s$,  emerge,
consider first  the limit $t_0\to \infty$. From  Eq. (\ref{hyper}), we
get after some algebra,
\begin{eqnarray}
C(t,t',\infty) & \sim & \alpha_n\,[t^{2n-1} + t'^{2n-1} - |t-t'|^{2n-1}]
,\ \frac{1}{2}< n< \frac{3}{2},\nonumber  \\
& \sim & \beta_n\,t_0^{2n-3}tt',\ n > \frac{3}{2},
\label{stat}
\end{eqnarray} 
where $\alpha_n=- \Gamma[n]\Gamma[1-2n]/\Gamma[1-n] >0$ and $\beta(n)=
[n^2-n+1  -  (n-2)/(2n-3)]/(2n-1)>0$.  To  calculate  the  persistence
properties  from  this  correlator,  we  note that  the  form  of  the
correlator for  $\frac{1}{2}< n<  \frac{3}{2}$ in Eq.  (\ref{stat}) is
precisely  that  of fbm  with  Hurst  exponent $H=n-\frac{1}{2}$.  The
corresponding persistence  exponent is known  exactly \cite{Krug2,fbm}
to  be $\theta_s =  1 -  H= \frac{3}{2}  - n  = (d+2-z)/2$,  from Eq.\
(\ref{n}),  provided   $z<d+2$.  For  $z>d+2$,   i.e.\  $n>3/2$,  Eq.\
(\ref{stat})        gives       the        normalized       correlator
$C(t,t',\infty)/\sqrt{C(t,t,\infty)C(t',t',\infty)}   =  1$,  implying
$\theta_s = 0$.

We now turn to the limit $t_0=0$. From Eq. (\ref{hyper}), we find that
the process is non-stationary in  time since the correlator depends on
both $t$  and $t'$ and not  just on the  difference $|t-t'|$. However,
using   the   standard   transformation\cite{review},   $X=x(t)/{\sqrt
{\langle x^2(t)\rangle}}$ and $T=\ln t$, one finds that $X(T)$ becomes
a Gaussian {\em  stationary} process in the ``new''  time variable $T$,
with a correlator $C_n(T)=\langle X(0)X(T)\rangle$,
\begin{equation}
C_n(T) = (2-{1\over {n}})\,e^{-T/2}\,F(1-n,1;1+n;e^{-T}),
\label{corr4}
\end{equation} 
where $F(a,b;c;z)$  is the standard  hypergeometric function\cite{GR}.
This form of  the correlator suggests that the  exponent $\theta_0$ is
nontrivial. To find  it one is confronted with  the following problem:
given  a   Gaussian  stationary  process  $X(T)$   with  a  prescribed
correlator  $C(T)=\langle X(0)X(T)\rangle$,  what  is the  probability
$P_0(T)$ that the process $X(T)$ does not cross zero up to time $T$?

For general correlator $C(T)$, $P_0(T)$ is hard to solve\cite{BL}. For
$n=1$, we  have, from  Eq. (\ref{corr4}), $C_1(T)=\exp(-T/2)$,  a pure
exponential.   This corresponds  to a  Markov process,  for  which the
exact result is known:  $P_0(T)\sim \exp (-T/2)\sim t^{-1/2}$ implying
$\theta_0(1)=1/2$.  For higher values  of $n$,  the process  $X(T)$ is
non-Markovian  and   hence  to  determine   $\theta_0(n)$  is  harder.
Fortunately  for  $n=2$  where $C_2(T)={3\over  {2}}\exp(-T/2)-{1\over
{2}}\exp(-3T/2)$,  the   exact  result  is   also  known:  $P_0(T)\sim
\exp(-T/4)\sim t^{-1/4}$\cite{TB1}.  For  general $n$, one can extract
some useful  information about the stochastic process  by studying the
short-time  properties  of  the  correlator $C_n(T)$.  Expanding  Eq.\
(\ref{corr4}) for small $T$ we get,
\begin{equation}
C_n(T)\simeq\cases{1-a_nT^{2n-1} & $1/2<n<3/2$,\cr 1+\frac{T^2}{4}\,
\ln T &$n=3/2$,\cr 1-\frac{(2n-1)}{8(2n-3)}\,T^2 &$n>3/2$,\cr}
\label{short}
\end{equation} 
where $a_n=\Gamma(n)\Gamma(2-2n)/\Gamma(1-n)$.   Thus for $n>3/2$, the
process is ``smooth'' with a  finite density of zero crossings that can
be derived using Rice's formula\cite{Rice},
\begin{equation}
\rho=\frac{\sqrt{-C_n''(0)}}{\pi} = 
\frac{1}{2\pi}\,\sqrt{\frac{2n-1}{2n-3}},
\label{density}
\end{equation}
where  ${C_n}''(0)$  is the  second  derivative  at  the origin.   For
$1/2<n<3/2$, the density  is infinite and the zeros  are not uniformly
distributed  but instead  form a  fractal set  with  fractal dimension
$d_f=n-1/2$\cite{BL}.  The case $n=3/2$  is marginal, and the integral
in Eq.\  (\ref{corr4}) can be  evaluated in closed  form, $C_{3/2}(T)=
\cosh(T/2)+\sinh^2(T/2)  \ln[\tanh(T/4)]$,   with  the  density  still
divergent   but  only   logarithmically.  A   physical   example  that
corresponds to this  marginal case $n=3/2$ is the  $2$-d DT model with
$z=4$. The exponent $\theta(n)$ diverges as $n\to 1/2$ from above. For
$n<1/2$  or equivalently  $z>d$  the continuum  equation  is not  well
defined and  one needs a non-zero  lattice constant in  space (for the
interface  problem)  or  time  (for  the stochastic  process  in  Eq.\
(\ref{st1})).

This transition at $n=n_c=3/2$ can  now be directly interpreted in the
interface context  using the relation  (\ref{n}).  Gaussian interfaces
with finite initial condition  undergoes a morphological transition at
$z_c=d+2$. For $z>z_c(d)$ ($n>3/2$),  the surface has a finite density
of  zero   crossings,  given  by  (\ref{n})   and  (\ref{density})  as
$\rho=(1/2\pi)\sqrt{(z-d)/(z-d-2)}$,  along  any  linear  cut  in  the
$d$-dimensional space. On the  other hand, for $z<z_c(d)$, the surface
crosses zero in a nonuniform way and the density is strictly infinite.
If it  crosses zero  once, it crosses  subsequently many  times before
making a long excursion.

This  transition is  reminiscent  of the  ``wrinkle'' transition  found
recently  by Toroczkai  et.  al.   \cite{Toro}  in $(1+1)$-dimensional
Gaussian    interfaces.     However     there    is    an    important
difference. Toroczkai  et.\ al.\ studied  the density of extrema  of a
Gaussian  $(1+1)$-dimensional surface  and found  that the  density of
extrema is finite for $z>5$ and infinite for $z<5$.  Their results can
be understood very simply within  our general framework by noting that
an extremum of the surface $h$ corresponds to ${\partial_x h}=0$. Thus
the density of  extrema of the surface $h$  corresponds to the density
of zeros of the derivative  process $\partial_xh$. One can then follow
our chain of arguments mapping  a Gaussian interface to the stochastic
process in Eq.\ (\ref{st1}) and  one finds that the derivative process
also maps to Eq.\ (\ref{st1}), but with $n= (z-d-1)/2$, i.e.\ $n$ gets
replaced   by   $(n+1)$  in   Eq.\   (\ref{n})   due   to  the   extra
derivative. Using $n_c=3/2$ again, we find that the transition for the
derivative   process  occurs   at   $z=z_c=d+4$.   This   is  thus   a
generalization of the $d=1$ result ($z_c=5$) of Ref. \cite{Toro}.

For $n>3/2$, where the process is  smooth, one can apply the IIA which
assumes  that  successive  intervals  between  zero-crossings  of  the
process $X(T)$  are statistically independent.  The IIA was  used very
successfully   for  rather  accurate   analytical  estimates   of  the
persistence  exponent  for  the  diffusion  equation\cite{diff,diff1}.
According to  this approximation, the exponent $\theta$  for a general
smooth Gaussian process  with correlator $C(T)$ is given  by the first
positive root of the following transcendental equation\cite{diff},
\begin{equation}
1+{
{2\theta}\over{\pi}}\int_0^{\infty}{\sin}^{-1}\left[C(T)\right]
e^{\theta T}dT={{2\rho}\over {\theta}},
\label{iia}
\end{equation}
where   $\rho={\sqrt  {-C''(0)}}/{\pi}$   is  the   density   of  zero
crossings. Applying  this formula to our problem,  with $C_n(T)$ given
by  Eq.\ (\ref{corr4})  and  $\rho$ by  Eq.\  (\ref{density}), we  can
obtain estimates for $\theta(n)$ for arbitrary $n>3/2$.

For example, putting $n=2$  in Eqs.\ (\ref{corr4}) and (\ref{iia}), we
get $\theta(2)=0.26466\ldots$ (IIA) which can be compared to the exact
result    $\theta(2)=1/4$.     Similarly    for   $n=3$,    we    find
$\theta(3)=0.22283\ldots   $(IIA),  to  be   compared  to   the  value
$\theta(3)\approx    0.231   \pm    .01$    obtained   by    numerical
simulation\cite{Cornell}.    The  limit   $n\to   \infty$  is   rather
interesting.  From  Eq.\  (\ref{corr4}), we  get,  $C_{\infty}(T)={\rm
{sech}} (T/2)$ which  happens to be also the  correlation function for
the two-dimensional diffusion equation\cite{diff,diff1}. For this, one
can  also   obtain  an  IIA   estimate,  $\theta(\infty)=0.1862\ldots$
(IIA)\cite{diff,diff1} which  is again  rather close to  the numerical
value  $\theta(\infty)\approx 0.1875\pm 0.0010$\cite{diff}.   Thus for
all $n>3/2$, the  analytical IIA estimates for $\theta(n)$  seem to be
rather accurate.

In the context of a stationary interface, the morphological transition
at $n=3/2$  (i.e. $z=d+2$) is associated with  the familiar transition
from  a   rough  to  a  super-rough  interface,   extracted  from  the
mean-square height difference  $C(x) = \langle [h(x)-h(0)]^2 \rangle$.
Using the stationary correlator, $\langle h_{\bf k} h_{-\bf k} \rangle
\sim  1/|{\bf k}|^z$ gives  $C(x) \sim  |x|^{z-d}$ for  $d<z<d+2$, and
$C(x) \sim x^2L^{z-d-2}$  for $z>d+2$, where the system  size $L$ acts
as a  regulator for the  small-$k$ divergence ($k_{min} \sim  1/L$) in
the latter case. This analysis also illustrates the difference between
finite and steady state ``initial'' points: if $h(0)$ is fixed, i.e. not
averaged over  the steady state  distribution, one obtains,  for large
$|x|$, $C(x) \sim  |x|^{z-d}$, independent of $L$, for  all $z>d$. The
same result  follows from Eq.\  (\ref{st1}) after the  replacements $x
\to h$, $t \to x$.

In summary, we  have shown that fluctuating interfaces  exhibit a form
of spatial persistence analogous to the temporal persistence exhibited
by  stochastic  processes.  For  Gaussian interfaces  the  analogy  is
precise  --  the spatial  fluctuations  in  the  stationary state  are
isomorphic to  the stochastic process  (\ref{st1}), with $n$  given by
Eq.\ (\ref{n}).  A {\em non-Gaussian} process for  which exact results
are  possible  is  the  $(1+1)$-dimensional KPZ  equation,  since  the
stationary probability  distribution of  the interface field  is given
by,       $P(\{h(x)\})\sim      \exp[-\int       (\partial_x      h)^2
dx]$\cite{Krug1}. Thus  the stationary  interface can be  described by
the  effective Langevin  equation, $\partial_xh  =\eta(x)$,  the $n=1$
version of  Eq.\ (\ref{st1}). From the  exact results $\theta(1)=1/2$,
and   $H=1/2$  we   immediately  obtain   $\theta_0=\theta_s=1/2$,  in
agreement  with  Ref.  \cite{DB}.  For  the  KPZ  equation  in  higher
dimensions, and other non-Gaussian interface models, the determination
of $\theta_0$, in particular\cite{manoj}, remains a challenge.

We thank M. Barma, G. Manoj, M. Marsili and C. Castellano for useful 
discussions.

\end{multicols}


\begin{references}
\bibitem{manoj} For $(2+1)$-dimensional KPZ interface, the relation 
$\theta_s=1-H$ has recently been verified by G. Manoj, private 
communication.
\end{references}
\end{document}